\documentstyle[preprint,aps]{revtex}
\tightenlines
\begin{document}

\title{Cosmic acceleration and a natural solution to the cosmological 
constant problem\footnote{gr-qc/9903005 v2, February 15, 2001}}

\author{\normalsize{Philip D. Mannheim} \\
\normalsize{Department of Physics,
University of Connecticut, Storrs, CT 06269} \\
\normalsize{mannheim@uconnvm.uconn.edu} \\}

\maketitle

\begin{abstract}
We trace the origin of the cosmological constant problem to the assumption that 
Newton's constant $G$ sets the scale for cosmology. And then we show that once 
this assumption is relaxed (so that the local $G$ as measured in a local 
Cavendish experiment is no longer to be associated with global cosmology), the 
very same cosmic acceleration which has served to make the cosmological 
constant problem so very severe instead then serves to provide us with its 
potential resolution. In addition, we present an alternate cosmology, one based 
on conformal gravity (a theory which explicitly possesses no fundamental $G$), 
and show that once given only that the sign of the vacuum energy density 
$\Lambda$ is explicitly the negative one associated with the spontaneous 
breakdown of the scale invariance of the conformal gravity theory (this 
actually being the choice of sign for $\Lambda$ which precisely leads to cosmic 
acceleration in the conformal theory), then that alone, no matter how big 
$\Lambda$ might actually be in magnitude, is sufficient to not only make its 
measurable contribution to current era cosmology naturally be of order one 
today, but to even do so in a way which is completely compatible with the 
recent high $z$ supernovae cosmology data. Cosmology can thus live with either 
a fundamental $G$ or with the large (and even potentially negative) $\Lambda$ 
associated with elementary particle physics phase transitions but not with 
both. In an appendix we distinguish between the thermodynamic free energy and 
the thermodynamic internal energy, with it being the former which determines 
cosmological phase transitions and the latter which is the source of the 
gravitational field. Then we show that, even if we make the standard ad hoc 
assumption that $\Lambda$ actually is quenched in standard gravity, vacuum 
energy density is nonetheless still found to dominate standard cosmology at the 
time of the phase transition which produced it. However, within conformal 
gravity no difficulty of this type is encountered.
\end{abstract}
\vfill\eject

\section{\bf ORIGIN OF THE COSMOLOGICAL CONSTANT PROBLEM}

The recent discovery \cite{Riess1998,Perlmutter1998} of a cosmic acceleration
has made the already extremely disturbing cosmological constant problem even 
more vexing than before. Specifically, a phenomenological fitting to the new 
high $z$ supernovae Hubble plot data using the standard Einstein-Friedmann 
cosmological evolution equations	
\begin{equation}
\dot{R}^2(t) +kc^2=\dot{R}^2(t)(\Omega_{M}(t)+\Omega_{\Lambda}(t))
\label{1}
\end{equation}
\begin{equation}
\Omega_{M}(t)+\Omega_{\Lambda}(t)+\Omega_{k}(t)=1
\label{1a2}
\end{equation}
\begin{equation}
\!\!\!\!\!q(t)=(n/2-1)\Omega_{M}(t)-\Omega_{\Lambda}(t)=
(n/2-1)(1+kc^2/\dot{R}^2(t)) - n\Omega_{\Lambda}(t)/2 
\label{1a3}
\end{equation}
where $\Omega_{M}(t)= 8\pi G\rho_{M}(t)/3c^2H^2(t)$ is due to ordinary matter 
(viz. matter for which $\rho_{M}(t)=A/R^n(t)$ where $A>0$ and $3\leq n \leq 4$), 
where $\Omega_{\Lambda}(t)= 8\pi G\Lambda/3cH^2(t)$ is due to a cosmological 
constant $\Lambda$ and where $\Omega_{k}(t)= -kc^2/\dot{R}^2(t)$
is due to the spatial 3-curvature $k$, has revealed that not only must the 
current era 
$\Omega_{\Lambda}(t_0)$ actually be non-zero today, it is even explicitly 
required to be of order one. Typically, the allowed parameter space compatible 
with the available data is found to be centered on the line 
$\Omega_{\Lambda}(t_0)=\Omega_{M}(t_0)+1/2$ or so, with $\Omega_{\Lambda}(t_0)$ 
being found to be limited to the range $(1/2,3/2)$ and 
(the presumed positive) $\Omega_{M}(t_0)$ 
to $(0,1)$, with the current ($n=3$)
era deceleration parameter 
$q(t_0)=(n/2-1)\Omega_{M}(t_0)-\Omega_{\Lambda}(t_0)$  
thus having to approximately lie within the $(-1/2,-1)$ interval. Thus, not 
only do we find that the universe is currently accelerating, but additionally 
we see that with there being no allowed $\Omega_{\Lambda}(t_0)=0$ solution at 
all (unless $\Omega_{M}(t_0)$ could somehow be allowed to go negative), the 
longstanding problem 
(see e.g. \cite{Weinberg1989,Ng1992} for some recent reviews) of trying to find 
some way by which $\Omega_{\Lambda}(t_0)$ could be quenched by many 
orders of magnitude from both its quantum gravity and particle physics
expectations (perhaps by making it vanish altogether) has now been replaced by 
the need to find a specific such mechanism which in practice (rather than just 
in principle) would explicitly put $\Omega_{\Lambda}(t_0)$ into this very narrow 
$(1/2,3/2)$ box.\footnote{In an appendix we show
how contrived the current situation actually needs to be.} Further, even 
independent of any quantum considerations, the 
new high $z$ data pose a problem for Eq. (\ref{1}) even when considered 
purely from the viewpoint of classical physics. Specifically, since the 
$\Omega_{\Lambda}(t)/\Omega_{M}(t)$ ratio evolves as $R^n(t)\sim T^{-n}(t)$, its 
current closeness to one entails that in the early universe this same ratio 
would have had to have been fantastically small, with the universe only being 
able to evolve into its
current state if this ratio had been extremely fine tuned in the early 
universe. Moreover, this particular fine tuning would have to be above and 
beyond that imposed by the flat inflationary universe 
model \cite{Guth1981} since inflation only constrains the sum of 
$\Omega_{M}(t)$ and $\Omega_{\Lambda}(t)$ to be one and does not fix their 
ratio. Thus at 
the present time neither inflationary nor quantum cosmology can readily 
accommodate the new high $z$ data at all. 
 
In order to try to diagnose the nature of the problem in as general a way as 
possible, we note that in any cosmology with a big bang, the early universe 
$\dot{R}(t=0)$ would have to be divergent (or at least be extremely large),
with Eq. (\ref{1}) then requiring the quantity $(\Omega_{M}(t=0)+
\Omega_{\Lambda}(t=0))$ to be equal to one no matter what the value of the 
spatial curvature $k$. Thus, given the radically different temporal behaviors 
of $\Omega_{M}(t)$ and $\Omega_{\Lambda}(t)$, in standard gravity no cosmology, 
flat or non-flat, could ever evolve into one in which 
$\Omega_{\Lambda}(t_0) \simeq \Omega_{M}(t_0) \simeq O(1)$ today without extreme 
fine tuning, to thus bring standard cosmology to a rather severe impasse which
challenges its viability. Further, if $\dot{R}(t=0)$ 
does start off divergent, it must diminish as the universe evolves, with the 
early universe thus decelerating. Since the current universe now appears to be 
accelerating, the fine tuning problem can be viewed as the need to adjust
parameters in such a way that the cosmology can exhibit diametrically opposite 
deceleration and acceleration behaviors in differing epochs. Since the big 
bang singularity itself derives from the fact that standard gravity is 
always attractive (since $G$ controls gravity on all distance scales 
including those much larger than the solar system one on which standard 
gravity was first established), while acceleration is more naturally 
associated with repulsion (cf. the fits to the supernovae data in which 
$\Omega_{M}(t_0)$ actually is taken to be negative), it is thus suggestive that 
we might be able to more readily balance the early 
and current universes if there were no initial singularity at 
all, and if cosmological gravity in fact got to be repulsive in all epochs,  
with the universe then expanding from some initially hot state characterized by 
the non-singular $\dot{R}(t=0)=0$ instead. To achieve this would thus appear to 
require the removal of $G$ from the fundamental gravitational action. 
Additionally, if $\dot{R}(t=0)$ were indeed to vanish, the initial value of 
$\Omega_{\Lambda}(t)$ would then be infinite no matter what, and would thus 
never require early universe fine tuning. Moreover, continuing in this same 
vein, we note that for an accelerating universe which actually accelerates 
indefinitely, then, no 
matter what may or may not have actually occurred in the early universe, at 
late times $\dot{R}(t)$ will eventually become arbitrarily large, with 
Eq. (\ref{1}) then requiring the quantity $(\Omega_{M}(t)+\Omega_{\Lambda}(t))$ 
to asymptote to one at late times, again independent of the value of $k$. 
However, because of their differing time behaviors, we see that in the 
late universe it would precisely be $\Omega_{\Lambda}$ which would then have to 
tend to one no matter what its early universe value. Thus at late enough times 
the cosmological constant problem would actually get solved, and in fact would 
get solved  by cosmology itself (i.e. no matter how big $\Lambda$ might 
actually be, in permanently accelerating universes there will eventually come 
a time in which the measurable consequence of $\Omega_{\Lambda}$ will be that it 
will make a contribution to the expansion of the universe which will be of 
order one). Thus even while the discovery of cosmic acceleration makes 
the cosmological constant problem more acute, nonetheless, its very existence 
also suggests a possible resolution of the issue. 

In order to see how we might be able to take advantage of this possibility, it
is very instructive to analyze \cite{Mannheim1998} de Sitter geometry in a 
purely kinematic way which requires no commitment to any particular dynamical 
equation of motion. Specifically, suppose we know only that a given geometry is 
de Sitter, i.e. that its Riemann tensor is given by 
\begin{equation}
R^{\lambda\rho\sigma\nu}
=\alpha(g^{\sigma \rho}g^{\lambda \nu}-g^{\nu \rho}g^{\lambda\sigma}).
\label{1b}
\end{equation}
For such a geometry contraction then yields the kinematic relation 
\begin{equation}
R^{\mu \nu} - 
g^{\mu \nu} R^{\sigma}_{\phantom {\sigma} \sigma}/2= 3\alpha g^{\mu \nu},
\label{1c}
\end{equation}
a relation which reduces to 
\begin{equation}
\dot{R}^2(t) +kc^2=\alpha c^2 R^2(t)
\label{1d}
\end{equation}
when evaluated 
in the Robertson-Walker coordinate system. On defining 
$\Omega_{\Lambda}(t) =\alpha c^2 R^2(t)/\dot{R}^2(t)$ we obtain 
$-q(t)=\Omega_{\Lambda}(t)=1+kc^2/\dot{R}^2(t)$, 
with $R(t)$, $q(t)$
and $\Omega_{\Lambda}(t)$ being found \cite{Mannheim1998} to given by
\begin{eqnarray}
R(t,\alpha<0,k<0)=(k/\alpha)^{1/2}sin((-\alpha)^{1/2}ct),
\nonumber \\
R(t,\alpha=0,k<0)=(-k)^{1/2}ct,
\nonumber \\
R(t,\alpha>0,k<0)=(-k/\alpha)^{1/2}sinh(\alpha^{1/2}ct),
\nonumber \\
R(t,\alpha>0,k=0)=R(t=0)exp(\alpha^{1/2}ct),
\nonumber \\
R(t,\alpha>0,k>0)=(k/\alpha)^{1/2}cosh(\alpha^{1/2}ct),
\nonumber \\
\label{1e}
\end{eqnarray}
\begin{eqnarray}  
\Omega_{\Lambda}(t,\alpha<0,k<0)=-q(t,\alpha<0,k<0)
=-tan^2((-\alpha)^{1/2}ct),
\nonumber \\ 
\Omega_{\Lambda}(t,\alpha=0,k<0)=-q(t,\alpha=0,k<0)=0, 
\nonumber \\
\Omega_{\Lambda}(t,\alpha>0,k<0)=-q(t,\alpha>0,k<0)=tanh^2(\alpha^{1/2}ct),
\nonumber \\
\Omega_{\Lambda}(t,\alpha>0,k=0)=-q(t,\alpha>0,k=0)=1,
\nonumber \\ 
\Omega_{\Lambda}(t,\alpha>0,k>0)=-q(t,\alpha>0,k>0)=coth^2(\alpha^{1/2}ct)
\label{1f}
\end{eqnarray}
in all of the various allowable cases. As we thus see, when the parameter 
$\alpha$ is positive, each associated solution corresponds to a permanently 
accelerating universe, and that in each such universe $\Omega_{\Lambda}(t)$ will 
eventually reach one no matter how big the parameter $\alpha$ might be, and 
independent in fact of whether or not $G$ even appears in the cosmological 
evolution equations at all. Moreover, while $\Omega_{\Lambda}(t,\alpha>0,k>0)$ 
will reach one at late times, quite remarkably, 
$\Omega_{\Lambda}(t,\alpha>0,k<0)$ will be bounded between 
zero and one at all times, no matter how large $\alpha$ might be. Thus 
unlike the unbounded $\alpha<0$ case, we see that when $\alpha$ is greater or 
equal to 
zero, $\Omega_{\Lambda}(t)$ is either bounded at all times or approaches a bound at 
late times. Late time $\alpha \geq 0$ de Sitter cosmologies will thus always
quench the contribution of a cosmological constant to cosmology no 
matter how large it may be, and thus the key task is to find a cosmology in 
which the current era is already sufficiently (but not too) late. 
Since $\Omega_M(t_0)$ is not zero today, 
the standard cosmology would not immediately appear to be an appropriate 
candidate, 
but, as we shall now see, this bounding mechanism will precisely be found to 
occur in conformal gravity, a theory which has recently been advanced as an 
alternative to standard gravity and its standard dark matter paradigm, a theory 
in which $G$ does not in fact set the scale for cosmology.

\vfill\eject

\section{\bf SOLUTION TO THE COSMOLOGICAL CONSTANT PROBLEM}

Conformal gravity (viz. gravity based on the locally conformal invariant 
Weyl action 
\begin{equation}
I_W=-\alpha_g \int d^4x (-g)^{1/2} C_{\lambda\mu\nu\kappa} 
C^{\lambda\mu\nu\kappa} 
\label{1f2}
\end{equation}
where $ C^{\lambda\mu\nu\kappa}$ is the conformal Weyl 
tensor and where $\alpha_g$ is a purely dimensionless gravitational coupling 
constant) has recently been advanced as a candidate gravitational theory 
because it has been found capable of addressing so many of the problems (such 
as dark matter) which currently afflict standard gravity 
(see e.g. \cite{Mannheim1998,Mannheim1999,Mannheim2000} \footnote{These 
references also give references to some of the large general literature on 
conformal gravity as well as provide further details of the work described 
in this paper.}). Its original motivation was the desire to give gravity a 
dimensionless coupling 
constant just like those associated with the three other fundamental 
interactions. And indeed \cite{Mannheim1990}, the local conformal symmetry 
invoked to do this then not only excludes the existence of any fundamental mass 
scales such as a fundamental cosmological constant, even after mass scales are 
induced by spontaneous breakdown of the conformal symmetry, the (still) 
traceless energy-momentum tensor then constrains any induced cosmological 
constant term to be of the same order of magnitude as all the other terms in 
$T^{\mu \nu}$, neither smaller nor larger. Thus, unlike standard gravity, 
precisely because of its additional symmetry, conformal gravity has a great deal 
of control over the cosmological constant. Essentially, with all mass scales - 
of gravity and particle physics both - being jointly generated by spontaneous 
breakdown of the scale symmetry, conformal gravity knows exactly where the zero 
of energy is. And, moreover, with the theory possessing no mass scale at all 
in the absence of 
symmetry breaking, the very lowering of the vacuum energy needed to actually 
spontaneously break the scale symmetry in the first place necessarily yields
an induced cosmological constant which is expressly negative. Conformal 
gravity thus has control not merely of the magnitude of the cosmological 
constant, it even has control of its sign, and it is the purpose of this paper 
to show that it is this very control which will provide for a completely natural 
accounting of the 
new high $z$ supernovae data.

The cosmology associated with conformal gravity was first presented in 
\cite{Mannheim1992} where it was shown to possess no flatness problem, to
thus release conformal cosmology from the need for the copious amounts of 
cosmological dark matter required of the standard theory. Subsequently 
\cite{Mannheim1996,Mannheim1998}, the cosmology was shown to also possess no 
horizon problem, no universe age problem, and, through negative spatial 
curvature, to naturally lead to cosmic repulsion.\footnote{In the same  way as 
a standard $k=0$ spatially flat inflationary cosmology with $\Lambda =0$ leads 
to $q(t_0)=1/2$, the analogous $\Lambda=0$ conformal cosmology presented in 
\cite{Mannheim1992} is a necessarily spatially open negatively curved 
$k<0$ one with a current era deceleration parameter which obeys $q(t_0)=0$ 
\cite{Mannheim1996}.
Thus, as had actually been noted well in advance of the recent supernovae data, 
even without a cosmological constant conformal cosmology already 
possesses a repulsion not present in the standard
theory.} To discuss conformal cosmology it is convenient to consider the 
conformal matter action 
\begin{equation}
I_M=-\hbar\int d^4x(-g)^{1/2}[S^\mu S_\mu/2+\lambda S^4
- S^2R^\mu_{\phantom{\mu}\mu}/12
+i\bar{\psi}\gamma^{\mu}(x)(\partial_\mu+\Gamma_\mu(x))\psi
-gS\bar{\psi}\psi]
\label{1g}
\end{equation}
for generic massless scalar and fermionic fields. For such 
an action, when the scalar field acquires a non-zero expectation value $S_0$, 
the entire energy-momentum tensor of the theory is found (for a perfect matter 
fluid $T^{\mu\nu}_{kin}$ of fermions) to take the form         
\begin{equation}
T^{\mu\nu}=T^{\mu\nu}_{kin}-\hbar S_0^2(R^{\mu\nu}-
g^{\mu\nu}R^\alpha_{\phantom{\alpha}\alpha}/2)/6            
-g^{\mu\nu}\hbar\lambda S_0^4;
\label{3}
\end{equation}
with the complete solution to the scalar field, fermionic field, and 
gravitational field equations of motion in a background Robertson-Walker 
geometry (viz. a geometry in which the Weyl tensor vanishes) then reducing to 
just one relevant equation, viz.
\begin{equation}
T^{\mu\nu}=0,
\label{1h}
\end{equation}
a remarkably simple condition which immediately fixes the zero of energy. We 
thus see that the evolution equation of conformal cosmology looks identical to 
that of standard gravity save only that the quantity $-\hbar S_0^2 /12$ has 
replaced the familiar $c^3/16 \pi G$. This change in sign compared with 
standard gravity leads to a cosmology in which gravity is globally repulsive 
rather than attractive, even while local solar system gravity remains 
attractive (and completely standard \cite{Mannheim1994}) in the conformal 
theory.\footnote{The sign of the gravity generated by local 
gravitational inhomogeneities is fixed \cite{Mannheim1994} by the sign of the 
coupling constant $\alpha_g$ in the Weyl action $I_W$ of Eq. (\ref{1f2}), a 
quantity which simply makes no contribution at all 
in highly symmetric cosmologically relevant geometries 
where the Weyl tensor vanishes, with the signs of local and global gravity thus 
being totally decoupled in the conformal theory, to thereby allow inhomogeneous 
locally attractive Cavendish gravity and homogeneous globally repulsive 
cosmological gravity to coexist.} Because of this change in sign, conformal 
cosmology thus has no 
initial singularity (i.e. it expands from 
a finite minimum radius), and is thus precisely released from the standard big 
bang model constraints described earlier. Similarly, because of this change in 
sign the contribution of $\rho_{M}(t)$ to the expansion of the universe is 
now effectively repulsive, to (heuristically) mesh with the phenomenological 
high $z$ 
data fits in which $\Omega_{M}(t)$ was allowed to go negative. Apart from a 
change in sign, we see that through $S_0$ there is also a change in the 
strength of gravity compared to the standard theory. It is this feature which 
will now enable us to provide a complete accounting of the high $z$ data.

Given the equation of motion $T^{\mu \nu}=0$, the ensuing conformal cosmology 
evolution equation is then found to take the form (on setting 
$\Lambda=\hbar\lambda S^4_0$)   
\begin{eqnarray}
\dot{R}^2(t) +kc^2 =
-3\dot{R}^2(t)(\Omega_{M}(t)+
\Omega_{\Lambda}(t))/ 4 \pi S_0^2 L_{PL}^2
\equiv \dot{R}^2(t)(\bar{\Omega}_{M}(t)+
\bar{\Omega}_{\Lambda}(t)), 
\nonumber \\
\bar{\Omega}_{M}(t)+\bar{\Omega}_{\Lambda}(t)+\Omega_{k}(t)=1,
\label{4}
\end{eqnarray}
with the deceleration parameter now being given as
\begin{equation}
q(t)=(n/2-1)\bar{\Omega}_{M}(t)-\bar{\Omega}_{\Lambda}(t)=
(n/2-1)(1+kc^2/\dot{R}^2(t)) - n\bar{\Omega}_{\Lambda}(t)/2.
\label{4a}
\end{equation}
As we see, Eq. (\ref{4}) is remarkably similar in form to Eq. (\ref{1}), with 
conformal cosmology thus only containing familiar ingredients. As an alternate 
cosmology then, conformal gravity thus gets about as close to standard gravity 
as it would appear possible for an alternative to get while nonetheless still 
being 
different. Moreover, even though that had not been its intent, because of this 
similarity, we see that phenomenological fits in which $\Omega_{M}(t)$ and 
$\Omega_{\Lambda}(t)$ are allowed to vary freely in Eq. (\ref{1}) are thus also 
in fact phenomenological fits to Eq. (\ref{4}), with the various $\Omega(t)$ 
simply being replaced by their barred counterparts. 

In order to see whether conformal gravity can thus fit into the relevant 
$ \bar{\Omega}_{\Lambda}(t_0) = \bar{\Omega}_{M}(t_0)+1/2$ window (a window 
which we note includes 
$\bar{\Omega}_{M}(t_0)=0$, $ \bar{\Omega}_{\Lambda}(t_0)=1/2$), it is thus 
necessary to analyze the solutions to Eq. (\ref{4}). Before 
doing this, however, we note that unlike the situation in standard gravity 
where the sign of $\Omega_{\Lambda}(t)$ is not a priori known (its
sign is only ascertained via the supernovae fitting itself), in conformal 
gravity we see that since, as noted earlier, the sign of $\Lambda$ is 
necessarily negative in the conformal theory, the cosmic repulsion associated 
with the negative effective gravitational coupling $-\hbar S_0^2 /12$ then 
entails that $\bar{\Omega}_{\Lambda}(t)$ itself is uniquely positive. And, 
indeed, not only is this sign unambiguously known in the conformal theory,  
according to Eq. (\ref{4a}) it immediately leads to none other than cosmic 
acceleration. Moreover, in conformal gravity not only is the sign of 
$\bar{\Omega}_{\Lambda}(t)$ known a priori, so also is the sign of $k$, with 
it having been shown \cite{Mannheim2000} that at temperatures above all 
elementary particle physics phase transitions, the condition 
$T^{\mu\nu}_{kin}=0$ (which is what Eqs. (\ref{3}) and (\ref{1h}) reduce 
to when $S_0=0$) is only satisfiable non-trivially when $k$ is negative, 
with the positive energy density of ordinary matter then being identically 
canceled by the negative gravitational energy density associated with 
negative spatial curvature. Beyond this theoretical argument, actual 
observational support for $k$ being negative in conformal gravity has even 
been independently obtained from a recent study of galactic rotation curves 
\cite{Mannheim1997}, where an effect due to a global $k<0$ cosmology on local 
galactic dynamics was identified, one which turns out to provide for a complete 
accounting of rotation curve systematics without the need to ever 
introduce any of the galactic dark matter required in standard gravity. 
Thus in the following we shall 
explicitly take $k$ to be negative in the conformal theory while noting 
immediately that according to Eq. (\ref{4a}) such negative $k$ not only also 
leads to cosmic acceleration, but with its associated $\Omega_k(t)$ 
contribution being positive, Eq. (\ref{4}) then even enables the necessarily 
positive $\bar{\Omega}_{\Lambda}(t)$ to be less than one. Given 
such structure in the conformal theory we shall thus now turn to the explicit 
solutions to the conformal theory to see whether the theory is 
capable of not only giving some possible  
cosmic acceleration but whether it can even give just the amount 
detected.           

Explicit solutions to the conformal theory are readily obtained 
\cite{Mannheim1998}, and they can be 
classified according to the signs of $\lambda$ and $k$. (For completeness and 
comparison purposes in the following we explore the behavior of the theory for 
all possible allowable signs of $\lambda$ and $k$ rather than for just our 
preferred negative ones.) In the simpler to treat high temperature era where 
$\rho_{M}(t)=A/R^4=\sigma T^4$ the complete family of 
possible solutions is given as 
\begin{eqnarray}
R^2(t,\alpha<0,k<0)=k(1-\beta)/2\alpha+
k\beta sin^2 (ct\sqrt{-\alpha})/\alpha,
\nonumber \\
R^2(t,\alpha=0,k<0)=-2A/k\hbar c S_0^2-kc^2t^2,
\nonumber \\
R^2(t,\alpha>0,k<0)= -k(\beta-1)/2\alpha
-k\beta sinh^2 (\alpha^{1/2} ct)/\alpha,
\nonumber \\
R^2(t,\alpha > 0,k=0)=(-A/\hbar\lambda c S_0^4)^{1/2}
cosh(2\alpha^{1/2}ct),
\nonumber \\
R^2(t,\alpha > 0,k>0)=k(1+\beta)/2\alpha+
k\beta sinh^2 (\alpha^{1/2} ct)/\alpha,
\label{5}
\end{eqnarray}
where we have introduced the parameters $\alpha =-2\lambda S_0^2$, 
and $\beta =(1- 16A\lambda/k^2\hbar c)^{1/2}$. Similarly the associated 
deceleration parameters take the form
\begin{eqnarray}
q(\alpha<0,k<0)=
tan^2(ct\sqrt{-\alpha})
-2(1-\beta)cos(2ct\sqrt{-\alpha})/
\beta sin^2(2ct\sqrt{-\alpha}),
\nonumber \\
q(\alpha=0,k<0)=-2A/k^2\hbar c^3 S_0^2t^2,
\nonumber \\
q(\alpha>0,k<0)=
-tanh^2(\alpha^{1/2}ct)
+2(1-\beta)cosh(2\alpha^{1/2}ct)/
\beta sinh^2(2\alpha^{1/2}ct),
\nonumber \\
q(\alpha>0,k=0)=-1-2/sinh^2(2\alpha^{1/2}ct),
\nonumber \\
\!\!\!\!
q(\alpha>0,k>0)=-coth^2(\alpha^{1/2}ct)
-2(1-\beta)cosh(2\alpha^{1/2}ct)/
\beta sinh^2(2\alpha^{1/2}ct).
\label{6}
\end{eqnarray}
Now while Eq. (\ref{5}) yields a variety of temporal behaviors for $R(t)$, it 
is of great interest to note that every single one of them begins 
with $\dot{R}(t=0)$ being zero (rather than infinite) just as desired above; 
and that when $\lambda$ is precisely our preferred negative value (viz.
$\alpha>0$) each associated such solution corresponds to 
a universe which permanently expands (only the 
$\lambda >0$ solution recollapses, with conformal cosmology thus 
correlating the long time behavior of $R(t)$ with the sign of $\lambda$ 
rather than with that of $k$). We thus need to determine the degree to 
which the permanently expanding universes have by now already become 
permanently accelerating.

To this end we note first from Eq. (\ref{6}) that with $\beta$ being greater 
than one when $\lambda$ is negative, both the $\alpha >0,~k<0$ and the 
$\alpha >0,~k=0$ cosmologies are in fact permanently accelerating ones no 
matter what the values of their parameters. To explore the degree to which 
they have by now already become asymptotic, as well as to
determine the acceleration properties of the $\alpha >0,~k>0$ cosmology, we 
note that since each of the solutions given in Eq. (\ref{5})
has a non-zero minimum radius, each associated $\alpha >0$ cosmology has some 
very large but finite maximum temperature $T_{max}$ given by
\begin{eqnarray}
T_{max}^2(\alpha>0,k<0)/T^2(t,\alpha>0,k<0)= 1
+2\beta sinh^2 (\alpha^{1/2} ct)/(\beta-1),
\nonumber \\
T_{max}^2(\alpha > 0,k=0)/T^2(t,\alpha>0,k=0)=
cosh(2\alpha^{1/2}ct),
\nonumber \\
T_{max}^2(\alpha > 0,k>0)/T^2(t,\alpha>0,k>0)=
1+2\beta sinh^2 (\alpha^{1/2} ct)/(\beta+1),
\label{7}
\end{eqnarray}
with all the permanently expanding ones thus necessarily being below their 
maximum temperatures today, and actually being way below once given enough 
time. To obtain further insight into these solutions it is convenient to 
introduce an effective temperature according to $-c\hbar \lambda S^4_0
=\sigma T_V^4$. In terms of this $T_V$ we then find that in all the 
$\lambda<0$ cosmologies the energy density terms take the form
\begin{eqnarray}
\bar{\Omega}_{\Lambda}(t)=(1-T^2/T_{max}^2)^{-1}(1+T^2T_{max}^2/T_V^4)^{-1},
\nonumber \\
\bar{\Omega}_M(t)=-(T^4/T_V^4)\bar{\Omega}_{\Lambda}(t),
\label{8}
\end{eqnarray}
where $(\beta-1)/(\beta+1)=T_V^4/T_{max}^4$ for the $k<0$ case, and where 
$(\beta-1)/(\beta+1)=T_{max}^4/T_V^4$ for the $k>0$ case. With $\beta$ being
greater than one, we find that for the $k>0$ case $T_V$ is greater than 
$T_{max}$, for $k=0$ $T_V$ is equal to $T_{max}$, and for $k<0$ $T_V$ is less 
than $T_{max}$, with the energy in curvature (viz. the energy in the 
gravitational field itself) thus making a direct contribution to the maximum 
temperature of the universe. Hence, simply because $T_{max}$ is 
overwhelmingly larger than the current temperature  $T(t_0)$ 
(i.e. simply because the universe has been expanding and cooling for such 
a long time now), we see, that without any fine tuning at all, 
in both the $k>0$ and $k=0$ cases (i.e. cases where $T_V \geq T_{max} 
\gg T(t_0)$) the quantity $\bar{\Omega}_{\Lambda}(t_0)$ must 
already be at its asymptotic limit of one today, that $\bar{\Omega}_M(t_0)$ 
must be completely suppressed, and that $q(t_0)$ must be equal to minus one. 

For the $k<0$ case however (the only $\alpha>0$ case where $T_V$ is less than 
$T_{max}$, with a large $T_V$ thus entailing an even larger $T_{max}$) 
an even more interesting and relevant outcome is possible. Specifically, 
since in this case 
the quantity $(1+T^2T_{max}^2/T_V^4)^{-1}$ is always bounded between 
zero and one no matter what the relative magnitudes of $T_V$, $T_{max}$ and 
$T(t)$, the simple fact that $T_{max} \gg T(t_0)$ (something which must 
anyway be the case if $T_V \gg T(t_0)$) entails that rather than having 
already reached its asymptotic value of one, $\bar{\Omega}_{\Lambda}(t_0)$ 
instead need only be bounded by it. With the temporal evolution of 
the $\alpha>0$, $k<0$ cosmology in Eq. (\ref{7}) being given by
\begin{equation}
tanh^2 (\alpha^{1/2} ct)=(1-T^2/T_{max}^2)/(T_{max}^2T^2/T_V^4+1), 
\label{8a}
\end{equation}
we see, and again without any fine tuning at all, that simply since because 
$T_{max}$ is so much greater than $T(t_0)$, the current value of 
$\bar{\Omega}_{\Lambda}(t_0)$ is given 
by the nicely bounded $tanh^2(\alpha^{1/2}ct_0)$, that  
$\bar{\Omega}_M(t_0)$ is again completely suppressed in the current era, 
and that the curvature contribution is given by the non-negative 
$\Omega_{k}(t_0)=1-\bar{\Omega}_{\Lambda}(t_0)=
sech^2(\alpha^{1/2}ct_0) $ ($\Omega_k(t_0)$ is positive 
for negative $k$), 
with negative spatial curvature now being able to 
explicitly contribute to current era cosmological evolution. Moreover, in the 
$\alpha>0$, $k<0$ case, the bigger the spatial curvature contribution gets to 
be  the further $\bar{\Omega}_{\Lambda}(t_0)$ will lie 
below one, with it taking a value close to one half the closer to 
$T_V^2/T_{max}$ the current 
temperature gets to be. 

Thus we see that in all three of the $\alpha >0$ cases the single simple 
requirement 
that $T_{max} \gg T(t_0)$ ensures 
that $\bar{\Omega}_M(t_0)$ is completely negligible at current temperatures 
(it can thus only be relevant in the early universe), with the current era Eq. 
(\ref{4}) then reducing to 
\begin{equation}
\dot{R}^2(t) +kc^2 
=\dot{R}^2(t)\bar{\Omega}_{\Lambda}(t)
=-\dot{R}^2(t)q(t), 
\label{8b}
\end{equation}
to thus not only yield as a current era conformal 
cosmology what in the standard theory could only possibly occur as a very late 
one, but to also yield one which enjoys all the nice purely kinematic 
properties of a de Sitter geometry which we identified above. Since studies of 
galaxy counts indicate that the purely visible matter contribution to 
$\Omega_{M}(t_0)$ is of order one (actually of order $10^{-3}$ or so in 
theories in which dark matter is not considered), it follows from Eq. (\ref{4}) 
that current era suppression of $\bar{\Omega}_{M}(t_0)$ will in fact be 
achieved if the conformal cosmology scale parameter $S_0$ is many orders of 
magnitude larger than $L^{-1}_{PL}$, a condition which is actually compatible 
with a large rather than a small $T_V$. Comparison with Eq. (\ref{1}) shows 
that current era $\lambda<0$ conformal cosmology looks exactly like a low mass 
standard model cosmology, except that instead of $\Omega_{M}(t_0)$ being 
negligibly small (something difficult to understand in the standard theory) it 
is $\bar{\Omega}_{M}(t_0)=-3\Omega_{M}(t_0)/4\pi S_0^2 L^2_{PL}$ which is 
negligibly small instead ($\Omega_{M}(t_0)$ itself need not actually be 
negligible in conformal gravity - rather, it is only the contribution of 
$\rho_{M}(t)$ to the evolution of the current universe which needs be small). 
Thus, to conclude we see that when $\lambda$ is negative, conformal cosmology 
automatically leads us to $\bar{\Omega}_{M}(t_0)=0$ and to 
$0 \leq \bar{\Omega}_{\Lambda}(t_0) \leq 1$, with 
$\bar{\Omega}_{\Lambda}(t_0)$ being smaller the more important 
the contribution $\Omega_{k}(t_0)$ of the spatial 
curvature of the universe to its expansion gets 
to be. Moreover, with it being the case that, as had been noted earlier, 
$k$ actually 
is negative in the conformal theory, we see that conformal 
gravity thus leads us directly to $\bar{\Omega}_{M}(t_0)=0$, 
$\bar{\Omega}_{\Lambda}(t_0)=tanh^2(\alpha^{1/2}ct_0)$, i.e. precisely right 
into the 
phenomenological region favored by the new high $z$ data.

As regards our treatment of the cosmological constant, it is important to 
stress that there is a big distinction between trying to make $\Lambda$ 
itself small (the standard way to try to address the cosmological constant 
problem) and trying to make its current contribution ($\Omega_{\Lambda}(t_0)$ or 
$\bar{\Omega}_{\Lambda}(t_0)$) to observational cosmology be small, with this 
latter 
possibility being all that is required by actual observational information. 
Moreover, independent of whether superstring quantum gravity is or is 
not capable of quenching a Planck density cosmological constant, spontaneous 
breakdown effects such as those associated with a Goldstone boson pion or with 
massive intermediate vector bosons are clearly very much in evidence in current 
era particle physics experiments, and thus not quenched apparently. Hence all
the evidence of particle physics is that its contribution to $\Lambda$ should 
in fact be large rather than small today, with the essence of our work here 
being that even in such a case $\bar{\Omega}_{\Lambda}(t_0)$ can nonetheless 
still be 
small today. Indeed, as our model independent study of de Sitter geometry 
showed, cosmic acceleration will cause the standard gravity 
$\Omega_{\Lambda}(t)$ and 
the conformal gravity $\bar{\Omega}_{\Lambda}(t)$ to both become small at late 
enough 
times no matter how large $\Lambda$ might be. Moreover, the $G$ independent 
ratio $\Omega_{M}(t)/ \Omega_{\Lambda}(t)
=\bar{\Omega}_{M}(t)/\bar{\Omega}_{\Lambda}(t)
=-T^4/T_V^4$ will also become very small at late times no matter what value of 
$G$ might be measured in a local Cavendish experiment. With the overall 
normalization of the contribution of $\rho_{M}(t)$ to cosmology being the only 
place where $G$ could possibly play any role cosmologically, we see that the 
standard gravity fine tuning problem associated with having $\Omega_{M}(t_0) 
\simeq \Omega_{\Lambda}(t_0)$ today can be viewed as being not so much one of 
trying 
to understand why it is $\Omega_{\Lambda}(t_0)$ which is of order one after 15 
or so 
billion years, but rather of trying to explain why the matter density 
contribution to cosmology should be of order one after that much time rather 
than a factor $T^4/T_V^4$ smaller. Since this latter problem is readily 
resolved if $G$ does not in fact control cosmology, but if cosmology is instead 
controlled by some altogether smaller scale such as $-1/S_0^2$ (indeed the 
essence of conformal cosmology is that the larger $S_0$, i.e. the larger rather 
than the smaller $\Lambda$, the faster $\bar{\Omega}_{M}(t)$ decouples from 
cosmology), we see that the origin of the entire cosmological constant problem 
can directly be traced to the assumption that gravity is controlled by Newton's 
constant $G$ on each and every distance scale. The author wishes to thank Drs. 
G. V. Dunne, B. R. Holstein, V. A. Miransky and M. Sher for helpful comments. 
This work has been supported 
in part by the Department of Energy under grant No. DE-FG02-92ER40716.00.

\vfill\eject

\section{\bf APPENDIX -  DIFFICULTIES IN QUENCHING $\Lambda$}

Beyond the fact that a quenching of the cosmological constant down from its 
quantum gravity and particle physics expectations has yet to be achieved for
standard gravity, it is important to note how contrived such a quenching would 
anyway need to be even it were to be achieved. Specifically, suppose that some 
explicit quenching does take place in the very early Planck temperature 
dominated standard model quantum gravity universe (some candidate mechanisms 
which might be able to do this are discussed in 
\cite{Weinberg1989,Ng1992}), so that at the end of that era (or at the end of a 
subsequent but still early inflationary universe era) $\Lambda$ then takes some 
specific and particular value (this value might even be due to a fundamental or 
anthropically induced cosmological constant which simply appears in the 
fundamental gravitational action as an a 
priori fundamental constant). As the ensuing universe then expands and cools it 
will potentially go through a whole sequence of elementary particle 
physics phase transitions, in each one of which the vacuum energy would be 
lowered (this being the definition of a phase transition). The residual 
$\Lambda$ from the early universe quenching would then have to be such that it 
would just almost (but not quite completely, now given the new supernovae data) 
cancel the net drop in vacuum energy due to all the particle physics phase 
transitions (transitions that would occur only after the early universe 
quenching had already taken 
place - unless each such phase transition is to be accompanied by its own 
quenching that is), so that just today, i.e. conveniently just 
for our own particular epoch, $\Omega_{\Lambda}(t_0)$ would then be of order 
one. Difficult as this is to even conceive of let alone demonstrate in an 
explicit dynamical model, we note, however, that even if such a delicate 
current era balancing were to actually take place, the resulting universe 
would then be one 
in which there might not necessarily have been such delicate cancellations 
in epochs prior to the current one. Thus in earlier epochs there could well 
have been substantial 
cosmological constant contributions, to thus potentially give the universe a 
history and cosmology very different from the one conventionally considered. 

To see how an undesired scenario such as this might actually occur, it is 
instructive to consider the standard elementary particle physics 
electroweak symmetry breaking phase transition, a transition 
which purely for illustrative purposes we initially model
by a Ginzburg-Landau theory with free energy effective potential
\begin{equation} 
V(\phi, T)=\lambda\phi^4/24 -\mu^2(T)\phi^2
\label{A1}
\end{equation}
where $\phi$ is the relevant order parameter and where $\mu^2(T)$ is 
given \cite{Weinberg1974} by a typical form such as 
$\mu^2(T)=\lambda(T_V^2-T^2)/12$. 
When the temperature $T$ is less than the transition 
temperature $T_V$, the potential $V(\phi, T)$ possesses a non-trivial minimum 
at $\phi^2=T_V^2-T^2$ in which it takes the temperature dependent value  
\begin{equation} 
V_{min}(T<T_V)=-\lambda(T_V^2-T^2)^2/24,
\label{A2}
\end{equation}
a value which is expressly negative (with respect to the zero value which
$V_{min}(T>T_V)$ takes above the critical temperature), with the zero 
temperature $V_{min}(T=0)$ taking the value 
$-\lambda T_V^4/24$. Now while the phase diagram of the phase transition is 
described by the free energy $V(\phi, T)$, gravity itself couples to the 
internal 
energy 
density \cite{Donoghue1986}, 
i.e. to
\begin{equation} 
U(T)=V(\phi, T) -T dV(\phi, T)/dT,
\label{A2a}
\end{equation}
an energy density which evaluates to 
\begin{equation} 
U_{min}(T<T_V)=-\lambda(T_V^4+2T_V^2T^2-3T^4)/24
\label{A2a}
\end{equation}
at the minimum of the free energy.

To this energy density we must now add on a very specifically 
and uniquely chosen 
additional residual vacuum energy density from the early universe, viz.
\begin{equation} 
U_{res}=\lambda(T_V^4+2T_V^2T^2(t_0)-3T^4(t_0))/24+\sigma T^4(t_0)
\label{A3}
\end{equation}
so as to thus give us a total energy density
\begin{equation} 
U_{tot}(T<T_V)=
-\lambda (T^2-T^2(t_0))(2T_V^2-3T^2-3T^2(t_0))/24+
\sigma T^4(t_0) 
\label{A4}
\end{equation}
and a current era $U_{tot}(T(t_0))$ which would then precisely be of order the 
energy density in order matter (generically given as $\sigma T^4(t_0)$) 
today.\footnote{Why exactly the residual early universe $U_{res}$ should 
so explicitly and so specifically depend on the current temperature $T(t_0)$ so 
that just our 
particular epoch of observers would see only the quenched energy density 
$\sigma T^4(t_0)$ remains of course to be explained.}  However, as we 
immediately see, for somewhat earlier epochs where $T_V > T \gg T(t_0)$, the 
total vacuum energy density $U_{tot}(T<T_V)$ would then be of order 
$-\lambda T^2_VT^2$ and 
(with $\lambda$ being related to the standard Higgs and vector 
boson masses as $\lambda \simeq e^2 M_H^2/M_W^2$) would thus be substantially 
larger 
than the black body energy density at the same temperature $T$. Since the 
temperature dependence of a standard $k=0$ radiation era cosmology with 
non-zero $\Lambda$ is given by $\dot{T}^2/T^2=(8 \pi G/3c^2)(\sigma T^4
+c\Lambda)$, we see that our very ability to track current era cosmology all 
the way back to the early universe is contingent on demonstrating 
that vacuum terms such as $\lambda T^2_VT^2$ are never of relevance.  
It is thus not sufficient to simply find a mechanism which 
quenches the cosmological constant once (in some particular chosen epoch). 
Rather one needs renewed, temperature dependent, quenching each and every time 
there is phase 
transition or new contribution to the vacuum energy.\footnote{While $U_{res}$ 
can cancel the leading term in $U_{min}(T<T_V)$, because it is 
temperature 
independent $U_{res}$ leaves the non-leading terms untouched.} 

Now, as regards loop correction modifications to tree approximation based 
wisdom,
we recall (see \cite{Dolan1974} or the recent review of \cite{Sher1989}) 
that 
the occupied energy mode one loop correction to the tree 
approximation $V_{tree}(\phi)=m^2\phi^2/2+\lambda\phi^4/24$ potential 
yields  a 
net effective potential
\begin{equation} 
V(\phi)=m^2\phi^2/2+\lambda\phi^4/24+[M^4ln(M^4/m^4)
-3(M^2-2m^2/3)^2]/128\pi^2
+V_T(\phi) 
\label{A5}
\end{equation}
where $M^2=m^2+\lambda \phi^2/2$ and where 
\begin{equation} 
V_T(\phi)={k^4T^4 \over 2\pi^2}\int^{\infty}_0 
dx x^2 ln(1-e^{-(x^2+M^2/k^2T^2)^{1/2}}),
\label{A6}
\end{equation}
with the associated energy density then being given by
\begin{equation} 
U(T)=V(\phi)-TdV_T(\phi)/dT
\label{A6}
\end{equation}
in this case.\footnote{With the leading term in 
$V_T(\phi)$ being 
$-\pi^2k^4T^4/90$ at high temperatures, Eq. (\ref{A6}) then yields 
$U=\pi^2k^4T^4/30$, the standard spinless particle black body energy density.}
Since the temperature dependent $V_T(\phi)$ is exponentially suppressed 
and thus insignificant at low 
temperatures, this time we need to add on to $U_{min}$ an early universe 
residual  
\begin{equation} 
U_{res}=-V(\phi_0)+\sigma T^4(t_0)
 \label{A7}
\end{equation}
where 
\begin{equation} 
V(\phi_0)=m^2\phi_0^2/2+\lambda\phi_0^4/24+[M_0^4ln(M_0^4/m^4)
-3(M_0^2-2m^2/3)^2]/128\pi^2
 \label{A8}
\end{equation}
is the value of $V(\phi)$ at its low temperature minimum $\phi_0$ 
(here $M_0^2=m^2+\lambda \phi^2_0/2$), to give us a total
energy density 
\begin{equation} 
U_{tot}(T)=V(\phi)-TdV_T(\phi)/dT+U_{res},
 \label{A9}
\end{equation}
an energy density whose current era  
value at $\phi=\phi_0$ is then the requisite $\sigma T^4(t_0)$. However, now 
extrapolating $U_{tot}(T)$ back to the critical region where $V_T(\phi)$ no 
longer is suppressed, and noting that the one and higher 
multi loop corrections to the 
effective potential will collectively generate an entire power series 
of $(T_V^2-T^2)^n$ effective Ginzburg-Landau type terms with both the complete 
all order $V(\phi)$ and the complete all order 
$dV(\phi)dT=(dV(\phi)/d\phi)(d\phi/dT)$
then vanishing at the critical point, we see that $U_{res}$ will 
remain uncanceled there and thus dominate the expansion rate of the universe 
in the 
critical region. Since this $U_{res}$ will dominate over black body 
radiation until temperatures above the critical point, we see that its 
presence could potentially yield a substantial modification to the 
cosmological history of the universe in such critical regions. And while 
its presence would not affect standard big bang nucleosynthesis (which 
occurs at temperatures way below the last phase transition, temperatures at 
which 
$V_T(\phi)$ is suppressed), nonetheless our ability to extrapolate back beyond 
the electroweak scale would still be severely impaired.

While we thus see that any (still to be found) current era cosmological 
constant cancellation mechanism will still create problems for cosmology in 
other epochs, 
we conclude this paper by noting that the above described extrapolation problem 
is not in fact encountered in the conformal cosmology theory which we have 
presented in this work, since in the conformal case no residual $U_{res}$ is 
ever needed. Rather, the complete all order energy density 
$U_{min}(T<T_V)=[V(\phi)-TdV_T(\phi)/dT]|_{\phi=\phi_0}$ itself is the entire 
contribution 
associated 
with vacuum breaking, an energy density 
which 
then vanishes identically at the critical point, 
one which, even while it takes the very large unquenched value $V(\phi_0)$ at 
low 
temperatures, still leads to a nicely quenched current era 
$\bar{\Omega}_{\Lambda}(t_0)$ of 
order one. 

\vfill \eject


\begin{thebibliography}{99}

\bibitem{Riess1998} A. G. Riess et. al., Observational evidence from supernovae 
for an accelerating universe and a cosmological constant, 
Astronom. J. {\bf 116}, 1009 (1998).

\bibitem{Perlmutter1998} S. Perlmutter et. al., Measurements of $\Omega$ and 
$\Lambda$ from 42 high-redshift supernovae, Astrophys. J. {\bf 517}, 
565 (1999).

\bibitem{Weinberg1989} S. Weinberg, The cosmological constant  problem, 
Rev. Mod. Phys. {\bf 61}, 1 (1989).

\bibitem{Ng1992} Y. J. Ng, The cosmological constant  problem, 
Int. J. Mod. Phys. D {\bf 1}, 145 (1992).

\bibitem{Guth1981} A. H. Guth, Inflationary universe: A possible 
solution to the horizon and flatness problems, Phys. Rev. D {\bf 23}, 347 (1981).

\bibitem{Mannheim1998} P. D. Mannheim, Implications of 
cosmic repulsion for gravitational theory, Phys. Rev. D {\bf 58}, 103511 (1998).

\bibitem{Mannheim1999} P. D. Mannheim, Cosmic acceleration as the solution 
to the cosmological constant problem, astro-ph/9910093.

\bibitem{Mannheim2000} P. D. Mannheim, Attractive and repulsive gravity, 
Founds. Phys. {\bf 30}, 709 (2000).

\bibitem{Mannheim1990} P. D. Mannheim,  Conformal cosmology with no 
cosmological constant, Gen. Relativ. Gravit. 
{\bf 22}, 289 (1990).

\bibitem{Mannheim1992} P. D. Mannheim, Conformal gravity and the flatness 
problem, Astrophys. J. {\bf 391}, 
429 (1992).

\bibitem{Mannheim1996} P. D. Mannheim, Conformal cosmology and the age 
of the universe, astro-ph/9601071.

\bibitem{Mannheim1994} P. D. Mannheim and D. Kazanas, Newtonian limit of 
conformal gravity and the lack of necessity of the second order Poisson 
equation, Gen. Relativ. Gravit. {\bf 26}, 337 (1994).

\bibitem{Mannheim1997} P. D. Mannheim, Are galactic rotation curves really 
flat?, Astrophys. J. {\bf 479}, 659 (1997).

\bibitem{Weinberg1974} S. Weinberg, Gauge and global symmetries at high 
temperature, Phys. Rev. D {\bf 9}, 3357 (1974).

\bibitem{Donoghue1986} J. F. Donoghue, B. R. Holstein and R. W. Robinett, 
Gravitational coupling at finite temperature, Phys. Rev. D 
{\bf 34}, 1208 (1986).

\bibitem{Dolan1974} L. Dolan and R. Jackiw, Symmetry behavior at finite 
temperature, Phys. Rev. D {\bf 9,} 3320 (1974).

\bibitem{Sher1989} M. Sher, Electroweak Higgs potential and vacuum stability, 
Phys. Rept. {\bf 179}, 273 (1989).

\end{thebibliography}
\end{document}